# Software (Re-)Engineering with PSF

*Bob Diertens*

Programming Research Group, Faculty of Science, University of Amsterdam

*ABSTRACT*

This paper investigates the usefulness of PSF in software engineering and reengineering. PSF is based on ACP (Algebra of Communicating Processes) and as some architectural description languages are based on process algebra, we investigate whether PSF can be used at the software architecture level, but we also use PSF at lower abstract levels. As a case study we reengineer the compiler from the Toolkit of PSF.

## 1. Introduction

In this paper we investigate the usefulness of PSF (Process Specification Formalism) and its accompanying Toolkit in software engineering and software reengineering. This is motivated by a range of previous examples of the use of process algebra [2] in the area of architectural description languages (ADL's). We mention *Wright* [1] (based on CSP [11]), *Darwin* [12] (based on the $\pi$-calculus [15]), and *PADL* [6], which is inspired by *Wright* and *Darwin* and focuses on architectural styles. We do not limit our attention to software architecture, but apply PSF at other design levels as well.

PSF is based on ACP (Algebra of Communicating Processes) [3] and ASF (Algebraic Specification Formalism) [4]. A description of PSF can be found in [13], [14], [7], and [8]. It is supported by a toolkit that contains among other components a compiler and simulator. A simulation can be coupled to an animation [9],[1] which can either be made by hand or be automatically generated from the PSF specification [10].

In software engineering and reengineering it is common practice to decompose systems into components that communicate with each other. The main advantage of this decomposition is that maintainance can be done on smaller components that are easier to comprehend. To allow a number of components to communicate with each other a so-called coordination architecture will be required. In connection with PSF we will make use of the ToolBus [5] coordination architecture, a software application architecture developed at the CWI (Amsterdam) and the University of Amsterdam. It utilizes a scripting language based on process algebra to describe the communication between software tools. A ToolBus script describes a number of processes that can communicate with each other and of course with various tools existing outside the ToolBus. The role of the ToolBus when executing the script is to coordinate the various tools in order to perform some complex task. A language-dependent adapter that translates between the internal ToolBus data format and the data format used by the individual tools makes it possible to write every tool in the language best suited for the task(s) it has to perform.
For larger systems, such a script can become rather complex and for that reason quite difficult to test and debug. Specification of a script in PSF enables one to apply the analysis tools available for PSF on the specification of the script. Moreover, if one or more tools have been specified in PSF the script may also be analyzed in combination with PSF specifications of components of the whole system.

As a case study, we reengineer the PSF compiler. At the start of the reengineering process this compiler consists of several components run by a driver, which makes it a suitable candidate for ToolBus based

1. This coupling is done with the use of the ToolBus and the whole application is specified in PSF. One can consider this a proof of concept for the very thing we are trying to investigate in this paper.

coordination. First, we develop a PSF library of ToolBus internals. We give an example specification to show how to use this library, and turn this specification into a ToolBus application. Thereafter we provide a specification of the compiler, from which we derive a specification of the compiler as a ToolBus application. We then turn the compiler into a real ToolBus application. A specification of the architecture for this (reengineered) compiler is extracted from its specification. Using this architectural specification, we then build a parallel version of the compiler, while reusing specifications and implementations for components of the compiler as it has already been configured as a ToolBus application.

## 2. Specification of the ToolBus library

This section presents a specification of a library of interfaces for PSF which can be used as a basis for the specification of ToolBus applications. This specification does not cover all the facilities of the ToolBus, but just what is necessary for the project at hand.

### *2.1 Data*

First, a sort is defined for the data terms used in the tools. An abstraction is made from the actual data used by the tools.

```
data module ToolTypes
begin
    exports
    begin
        sorts
            Tterm
    end
end ToolTypes
```

Next, the sorts are introduced for the data terms and identifiers which will be used inside the ToolBus as well as for communication with the ToolBus.

```
data module ToolBusTypes
begin
    exports
    begin
        sorts
            TBterm,
            TBid
    end
end ToolBusTypes
```

The module ToolFunctions provides names for conversions between data terms used outside and inside the ToolBus.

```
data module ToolFunctions
begin
    exports
    begin
        functions
            tbterm : Tterm -> TBterm
            tterm : TBterm -> Tterm
    end
    imports
        ToolTypes,
        ToolBusTypes
    variables
        t : -> Tterm
    equations
    [’] tterm(tbterm(t)) = t
end ToolFunctions
```

The ToolBus has access to several functions operating on different types. Here only the operators for tests about equality and inequality of terms, will be needed. These are introduced in the module ToolBusFunctions.

```
data module ToolBusFunctions
```

```
begin
    exports
    begin
        functions
            equal : TBterm # TBterm -> BOOLEAN
    end
    imports
        ToolBusTypes,
        Booleans
    variables
        tb1 : -> TBterm
        tb2 : -> TBterm
    equations
    [’] equal(tb1, tb1) = true
    [’] not(equal(tb1, tb2)) = true
end ToolBusFunctions
```

### *2.2 Connecting tools to the ToolBus*

In Figure 1 two possible ways of connecting tools to the ToolBus are displayed. One way is to use a separate adapter and the other to have a builtin adapter. Tool1 communicates with its adapter over pipelines.[2]

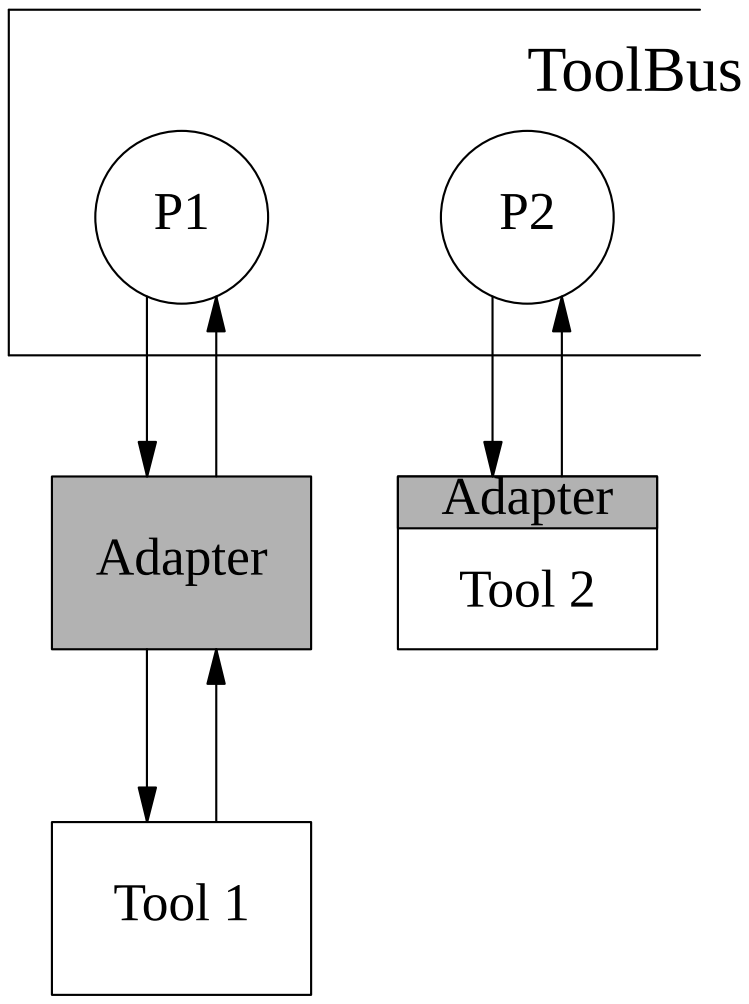

**Figure 1.** Model of tool and ToolBus interconnection

Next we define the primitives for communication between a tool and its adapter.

```
process module ToolAdapterPrimitives
begin
    exports
    begin
        atoms
            tooladapter-rec : Tterm
            tooladapter-snd : Tterm
    end
    imports
        ToolTypes
end ToolAdapterPrimitives
```

The primitives for communication between a tool and the ToolBus are fixed by the ToolBus design. At this stage these need to be formally defined in PSF, however. These primitives can be used for communication

2. In Unix systems, a pipeline is a means of communication between two processes.

between an adapter and the ToolBus as well, since the adapter logically takes the place of the tool it is supposed to connect to the ToolBus.

```
process module ToolToolBusPrimitives
begin
   exports
   begin
      atoms
         tooltb-snd : TBterm
         tooltb-rec : TBterm

         tooltb-snd-event : TBterm
         tooltb-rec-ack-event : TBterm
   end
   imports
      ToolBusTypes
end ToolToolBusPrimitives
```

Inside a ToolBus script a number of primitives may be used consisting of the actions for communication between ToolBus processes and their synchonous communication action, the actions used to communicate with the tools, and the action required to shutdown the ToolBus.

```
process module ToolBusPrimitives
begin
   exports
   begin
      atoms
         tb-snd-msg : TBterm # TBterm
         tb-rec-msg : TBterm # TBterm
         tb-comm-msg : TBterm # TBterm
         tb-snd-msg : TBterm # TBterm # TBterm
         tb-rec-msg : TBterm # TBterm # TBterm
         tb-comm-msg : TBterm # TBterm # TBterm

         tb-snd-eval : TBid # TBterm
         tb-rec-value : TBid # TBterm
         tb-snd-do : TBid # TBterm
         tb-rec-event : TBid # TBterm
         tb-snd-ack-event : TBid # TBterm

         tb-shutdown
   end
   imports
      ToolBusTypes
   communications
      tb-snd-msg(tb1, tb2) | tb-rec-msg(tb1, tb2) = tb-comm-msg(tb1, tb2)
         for tb1 in TBterm, tb2 in TBterm
      tb-snd-msg(tb1, tb2, tb3) | tb-rec-msg(tb1, tb2, tb3) = tb-comm-msg(tb1, tb2, tb3)
         for tb1 in TBterm, tb2 in TBterm, tb3 in TBterm
end ToolBusPrimitives
```

The ToolBus provides primitives allowing an arbitrary number of terms as parameters for communication between processes in the ToolBus. Here, the specification only covers the case of two and three term arguments for the primitives, because versions with more are usually not needed. In order to do better lists of terms have to be introduced, which is entirely possible in PSF but an unnececcary complication at this stage. The two-term version can be used with the first term as a ’to’ or ’from’ identifier and the second as a data argument. The three-term version can be used with the first term as ’from’, the second as ’to’, and the third as the actual data argument. If more arguments have to be passed, they can always be grouped into a single argument.

The module NewTool is a generic module with parameter Tool for connecting a tool to the ToolBus.

```
process module NewTool
begin
   parameters
      Tool
      begin
         processes
            Tool
      end Tool
```

```
        exports
        begin
           atoms
              tooltb-snd-value : TBid # TBterm
              tooltb-rec-eval : TBid # TBterm
              tooltb-rec-do : TBid # TBterm
              tooltb-snd-event : TBid # TBterm
              tooltb-rec-ack-event : TBid # TBterm
           processes
              TBProcess
           sets
              of atoms
                 TBProcess = {
                    tb-rec-value(tid, tb), tooltb-snd(tb),
                    tb-snd-eval(tid, tb), tb-snd-do(tid, tb),
                    tooltb-rec(tb), tb-rec-event(tid, tb),
                    tooltb-snd-event(tb), tb-snd-ack-event(tid, tb),
                    tooltb-rec-ack-event(tb)
                    | tid in TBid, tb in TBterm
                 }
        end
        imports
           ToolToolBusPrimitives,
           ToolBusPrimitives
        communications
           tooltb-snd(tb) | tb-rec-value(tid, tb) = tooltb-snd-value(tid, tb)
              for t in TBterm, tid in TBid
           tooltb-rec(tb) | tb-snd-eval(tid, tb) = tooltb-rec-eval(tid, tb)
              for t in TBterm, tid in TBid
           tooltb-rec(tb) | tb-snd-do(tid, tb) = tooltb-rec-do(tid, tb)
              for t in TBterm, tid in TBid
           tooltb-snd-event(tb) | tb-rec-event(tid, tb) = tooltb-snd-event(tid, tb)
              for t in TBterm, tid in TBid
           tooltb-rec-ack-event(tb) | tb-snd-ack-event(tid, tb) = tooltb-rec-ack-event(tid, tb)
              for tb in TBterm, tid in TBid
        definitions
           TBProcess = encaps(TBProcess, Tool)
     end NewTool
```

The process Tool accomplishes the connection between a process inside the ToolBus and a tool outside the ToolBus. The process TBProcess encapsulates the process Tool in order to enforce communications and thereby to prevent communications with other tools or processes. Note that TBProcess is used as the name of the main process and as the name of the encapsulation set. By doing so, they can both be renamed with a single renaming. This renaming is necessary if more than one tool is connected to the ToolBus (which is of course the whole point of the ToolBus).

The module NewToolAdapter is a generic module with parameters Tool and Adapter for connecting a tool and its adapter.

```
     process module NewToolAdapter
     begin
        parameters
           Tool
           begin
              atoms
                 tool-snd : Tterm
                 tool-rec : Tterm
              processes
                 Tool
           end Tool,
           Adapter
           begin
              processes
                 Adapter
           end Adapter
        exports
        begin
           atoms
              tooladapter-comm : Tterm
              adaptertool-comm : Tterm
           processes
```

```
            ToolAdapter
        sets
            of atoms
                ToolAdapter = {
                    tool-snd(t), tooladapter-rec(t),
                    tool-rec(t), tooladapter-snd(t)
                    | t in Tterm
                }
    end
    imports
        ToolAdapterPrimitives,
        ToolBusTypes
    communications
        tool-snd(t) | tooladapter-rec(t) = tooladapter-comm(t) for t in Tterm
        tool-rec(t) | tooladapter-snd(t) = adaptertool-comm(t) for t in Tterm
    definitions
        ToolAdapter = encaps(ToolAdapter, Adapter || Tool)
end NewToolAdapter
```

The process ToolAdapter puts an Adapter and a Tool in parallel and enforces communication between them with an encapsulation. In this case the main process and the encapsulation set have the same name once more, so that only one renaming is needed.

### *2.3 ToolBus instantiation*

The module NewToolBus is a generic module with parameter Application for instantiation of the ToolBus with an application.

```
process module NewToolBus
begin
    parameters
        Application
        begin
            processes
                Application
        end Application
    exports
    begin
        processes
            ToolBus
    end
    imports
        ToolBusPrimitives
    atoms
        application-shutdown
        tbc-shutdown
        tbc-app-shutdown
        TB-shutdown
        TB-app-shutdown
    processes
        ToolBus-Control
        Shutdown
    sets
        of atoms
            H = {
                tb-snd-msg(tb1, tb2), tb-rec-msg(tb1, tb2),
                tb-snd-msg(tb1, tb2, tb3), tb-rec-msg(tb1, tb2, tb3)
                | tb1 in TBterm, tb2 in TBterm, tb3 in TBterm
            }
            TB-H = {
                tb-shutdown, tbc-shutdown,
                tbc-app-shutdown, application-shutdown
            }
            P = { TB-shutdown, TB-app-shutdown }
    communications
        tb-shutdown | tbc-shutdown = TB-shutdown
        tbc-app-shutdown | application-shutdown = TB-app-shutdown
    definitions
        ToolBus =
            encaps(TB-H,
```

```
            prio(P > atoms,
               ToolBus-Control
            ||  disrupt(
                   encaps(H, Application),
                   Shutdown
               )
            )
         )
      ToolBus-Control = tbc-shutdown . tbc-app-shutdown
      Shutdown = application-shutdown
   end NewToolBus
```

A toolbus application can be described more clearly with `ToolBus = encaps(H, Application)`. The remaining code is needed to force a shutdown of all processes that otherwise would be left either running or in a state of deadlock after a ToolBus shutdown by the application. When an application needs to shutdown it performs an action `tb-shutdown` which will communicate with the action `tbc-shutdown` of the `ToolBus-Control` process, which then performs a `tbc-app-shutdown` that will communicate with `application-shutdown` of the `Shutdown` process enforcing a disrupt of the `Application` process.

In Figure 2 an overview is given of the import relations of the modules in the PSF ToolBus library. The module Booleans stems from a standard library of PSF.

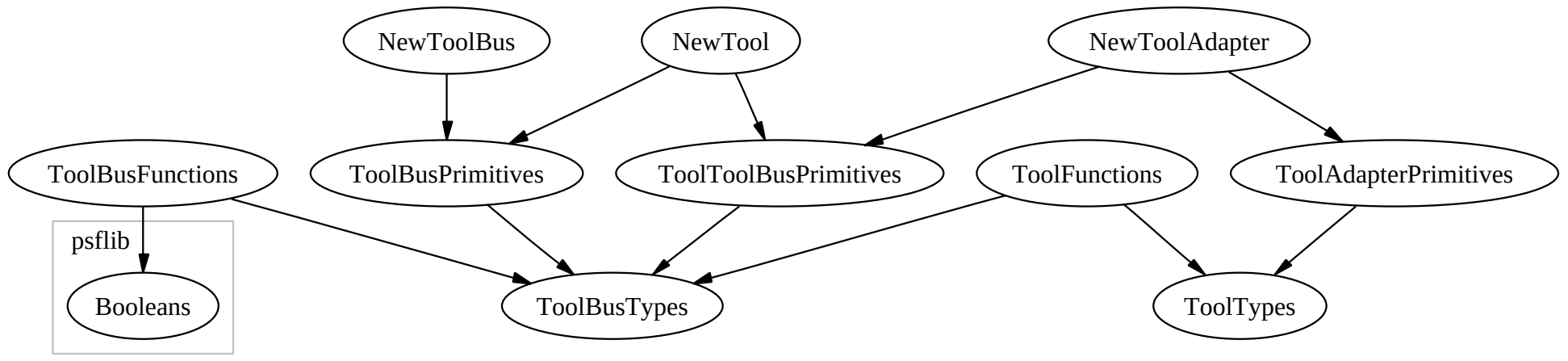

**Figure 2.** import graph of the ToolBus library

### *2.4 Example*

As an example of the use of the PSF ToolBus library, the specification is given of an application like the one shown in Figure 1. In this example, Tool1 can either send a 'message' to Tool2 and then wait for an acknowledgement from Tool2, or it can send a 'quit' after which the application will shutdown.

#### *2.4.1 Specification of the tools*

The first module defines the data that will be used.

```
data module Data
begin
   exports
   begin
      functions
         message : -> Tterm
         ack : -> Tterm
         quit : -> Tterm
   end
   imports
      ToolTypes
end Data
```

A specification of Tool1 and its adapter is then obtained.

```
process module Tool1
begin
   exports
   begin
      atoms
```

```
      snd : Tterm
      rec : Tterm
    processes
      Tool1
  end
  imports
    Data
  definitions
    Tool1 =
      (
        snd(message) .
        sum(d in Tterm, rec(d))
      +  snd(quit)
      ) . Tool1
end Tool1
```

```
process module AdapterTool1
begin
  exports
  begin
    processes
      AdapterTool1
  end
  imports
    ToolFunctions,
    ToolAdapterPrimitives,
    ToolToolBusPrimitives
  definitions
    AdapterTool1 =
      sum(d in Tterm,
        tooladapter-rec(d) .
        tooltb-snd-event(tbterm(d)) .
        sum(r in TBterm,
          tooltb-rec-ack-event(r) .
          tooladapter-snd(tterm(r))
        )
      ) . AdapterTool1
end AdapterTool1
```

Tool1 and its adapter are combined by importing NewToolAdapter and binding the parameters.

```
process module Tool1Adapter
begin
  imports
    NewToolAdapter {
      Tool bound by [
        tool-snd -> snd,
        tool-rec -> rec,
        Tool -> Tool1
      ] to Tool1
      Adapter bound by [
        Adapter -> AdapterTool1
      ] to AdapterTool1
      renamed by [
        ToolAdapter -> Tool1Adapter
      ]
    }
end Tool1Adapter
```

We specify Tool2

```
process module Tool2
begin
  exports
  begin
    processes
      Tool2
  end
  imports
    Data,
    ToolFunctions,
    ToolToolBusPrimitives
  definitions
```

```
      Tool2 =
         sum(d in TBterm,
            tooltb-rec(d) .
            tooltb-snd(tbterm(ack))
         ) . Tool2
end Tool2
```

### 2.4.2 *Specification of the ToolBus processes*

Some identifiers are defined in order to distinguish the messages sent between ToolBus processes themselves and between ToolBus processes and their accompanying tools. The lowercase identifiers (of type TBterm) are used with the actions `tb-snd-msg` and `tb-rec-msg`. The first argument of a message will always be the origin of the message, and the second argument will serve as its destination. Uppercase identifiers (of type TBid) are used as tool identifiers. Strictly speaking these are not necessary, since there can't be any communication with any other tool because of encapsulation. By using them, however, the actions for communication with a tool will have more similarity to the ones used in the ToolBus.

```
data module ID
begin
   exports
   begin
      functions
         T1 : -> TBid
         t1 : -> TBterm
         T2 : -> TBid
         t2 : -> TBterm
   end
   imports
      ToolBusTypes
end ID
```

For both tools a ToolBus process is defined. The specifications for these processes describe the protocol for communication between the tools.

```
process module PTool1
begin
   exports
   begin
      processes
         PTool1
   end
   imports
      Tool1Adapter,
      ID,
      ToolBusPrimitives,
      ToolBusFunctions
   processes
      PT1
   definitions
      PTool1 = Tool1Adapter || PT1
      PT1 =
         sum(d in TBterm,
            tb-rec-event(T1, d) .
            (
               [equal(d, tbterm(quit)) = true] ->
                  tb-shutdown
            +  [not(equal(d, tbterm(quit))) = true] -> (
                  tb-snd-msg(t1, t2, d) .
                  sum(r in TBterm,
                     tb-rec-msg(t2, t1, r) .
                     tb-snd-ack-event(T1, d)
                  )
               )
            )
         ) . PT1
end PTool1

process module PTool2
```

```
begin
   exports
   begin
      processes
         PTool2
   end
   imports
      Tool2,
      ID,
      ToolBusPrimitives
   processes
      PT2
   definitions
      PTool2 = Tool2 || PT2
      PT2 =
         sum(d in TBterm,
            tb-rec-msg(t1, t2, d) .
            tb-snd-eval(T2, d)
         ) .
         sum(r in TBterm,
            tb-rec-value(T2, r) .
            tb-snd-msg(t2, t1, r)
         ) . PT2
end PTool2
```

### *2.4.3 Specification of the ToolBus application*

The ToolBus processes are connected with the tools and together they constitute the process Run that merges the resulting two processes.

```
process module Tools
begin
   exports
   begin
      processes
         Run
   end
   imports
      NewTool {
         Tool bound by [
            Tool -> PTool1
         ] to PTool1
         renamed by [
            TBProcess -> XPTool1
         ]
      },
      NewTool {
         Tool bound by [
            Tool -> PTool2
         ] to PTool2
         renamed by [
            TBProcess -> XPTool2
         ]
      },
      ID,
      ToolBusFunctions
   definitions
      Run = XPTool1 || XPTool2
end Tools
```

At this stage renamings are necessary to be able to distinguish the two processes TBProcess.

The process Run is now transformed into a ToolBus application.

```
process module App
begin
   imports
      NewToolBus {
         Application bound by [
            Application -> Run
         ] to Tools
```

```
        }
    end App
```

The main process of this application is ToolBus. A generated animation is shown in Figure 3, in which AdapterTool1 just sent a message it had received from Tool1, to ToolBus process PT1.

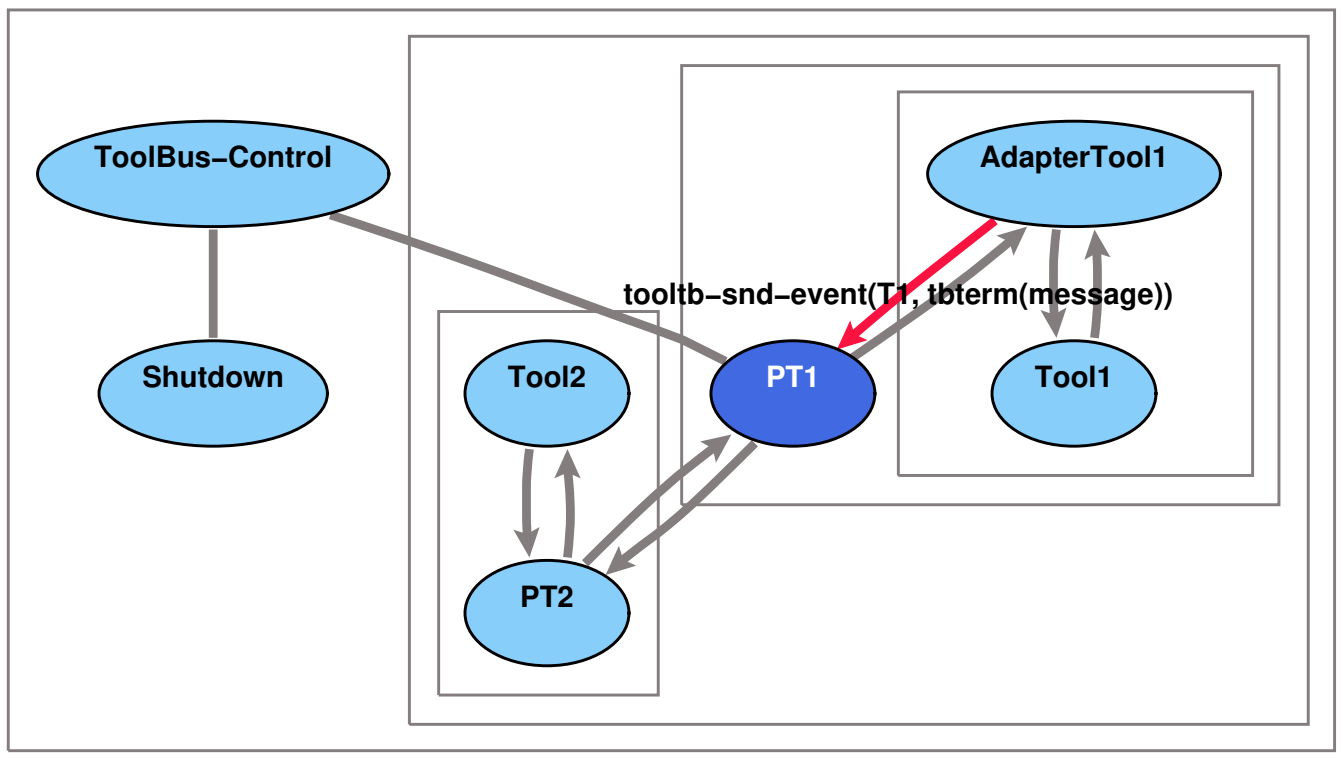

**Figure 3.** Animation of the ToolBus specification example

Each box represents an encapsulation of the processes inside the box, and a darker ellipse is a process which is enabled to perform an action in the given state.

### *2.4.4 Example as ToolBus application*

The application we have specified above has been built as an application consisting of three Tcl/Tk [16] programs (Tool1, its adapter, and Tool2), and a ToolBus script. A screendump of this application at work together with the viewer[3] of the ToolBus is shown in Figure 4.

---

3. With the viewer it is possible to step through the execution of the ToolBus script and view the variables of the individual processes inside the ToolBus.

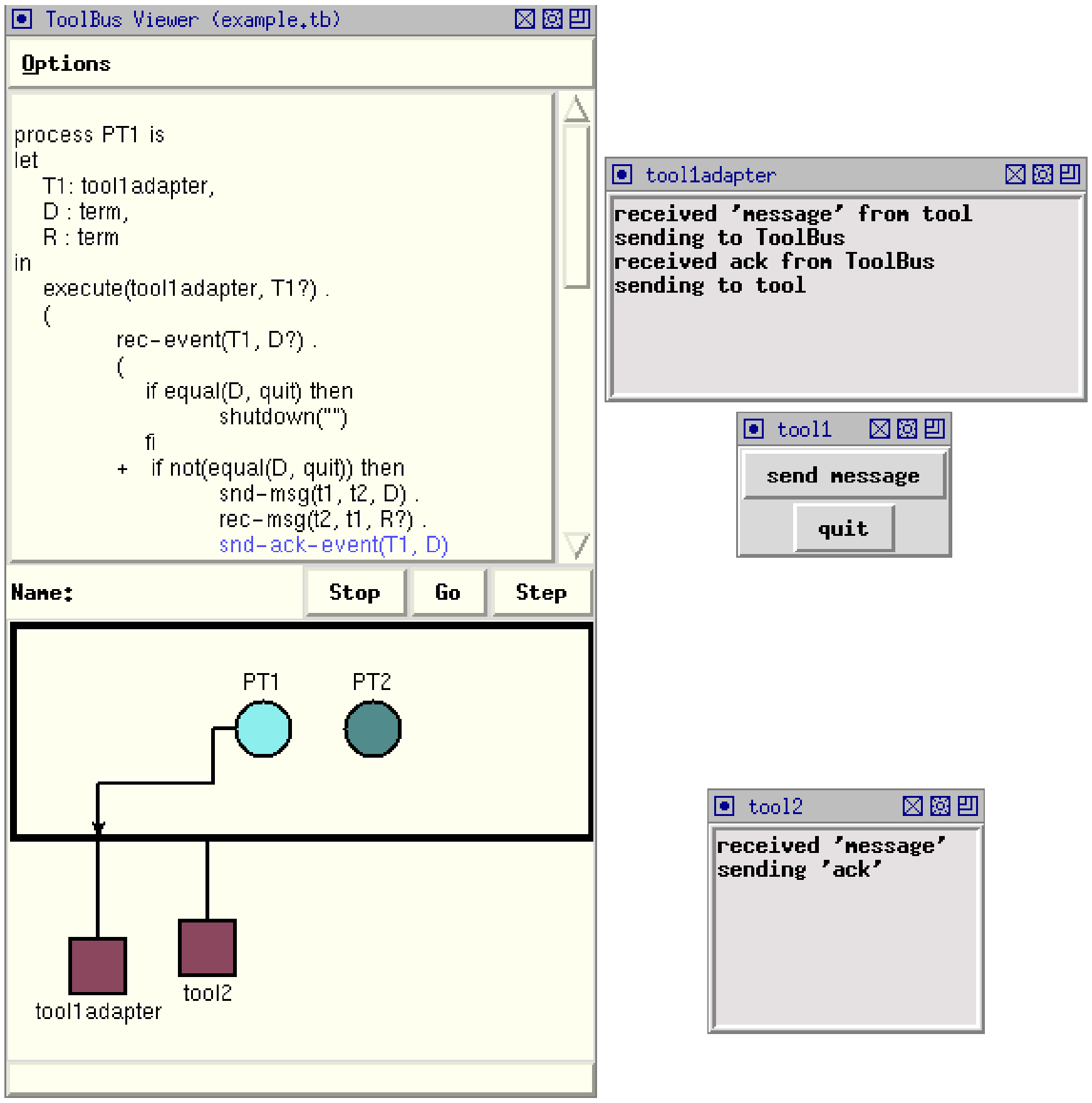

**Figure 4.** Screendump of the example as ToolBus application with viewer

The ToolBus script is shown below. The processes `PT1` and `PT2` closely resemble the processes `PTool1` and `PTool2` in our PSF specification. The `execute` actions in the ToolBus script correspond to starting of the adapter for Tool1 and starting of Tool2 in parallel with the processes `PT1` and `PT2` respectively.

```
process PT1 is
let
   T1: tool1adapter,
   D : term,
   R : term
in
   execute(tool1adapter, T1?) .
   (
      rec-event(T1, D?) .
      (
         if equal(D, quit) then
            shutdown("")
         fi
      +  if not(equal(D, quit)) then
            snd-msg(t1, t2, D) .
            rec-msg(t2, t1, R?) .
            snd-ack-event(T1, D)
         fi
      )
   ) * delta
endlet

process PT2 is
let
   T2: tool2,
   D : term,
   R : term
in
```

```
    execute(tool2, T2?) .
    (
      rec-msg(t1, t2, D?) .
      snd-eval(T2, eval(D)) .
      rec-value(T2, value(R?)) .
      snd-msg(t2, t1, R)
    ) * delta
  endlet

  tool tool1adapter is { command = "wish-adapter -script tool1adapter.tcl" }
  tool tool2 is { command = "wish-adapter -script tool2.tcl" }

  toolbus(PT1, PT2)
```

The actions `snd-eval` and `rec-value` differentiate from their equivalents in the PSF specification. The term `eval(D)` instead of just `D` is needed because the interpreter of evaluation requests that a tool receives from the ToolBus, calls a function with the name it finds as function in this term. We could have used any name instead of `eval` provided that Tool2 has got a function with that name.
Why the same scheme is needed by the ToolBus for `rec-value` is not known.

The processes in the ToolBus script use iteration and the processes in the PSF specification recursion. In PSF it is also possible to use iteration in this case, since the processes have no arguments to hold the current state. On the other hand, in PSF it is not possible to define variables for storing a global state, so when it is necessary to hold the current state, this must be done through the arguments of a process and be formalized via recursion.

The last line of the ToolBus script starts the processes `PT1` and `PT2` in parallel. Its equivalent in the PSF specification is the process `Run`.

## 3. Reengineering the PSF compiler

The PSF compiler is reengineered by developing a PSF specification for the compiler. From this specification we develop a second specification that makes use of the PSF library for the ToolBus, which will then be used for implementing a version of the compiler coordinated via the ToolBus.

### *3.1 Description of the compiler*

The PSF compiler translates a group of PSF modules to a tool interface language (TIL) that is suitable for tools to operate on. This compilation takes place in several phases. First each PSF module is parsed and converted to an MTIL (modular TIL) module. Then each MTIL module is normalized with as a result an ITIL (intermediate TIL) module. In this normalization step all imports are resolved by combining the MTIL module with the ITIL modules that correspond to the imported modules. The resulting module no longer depends on any imports. The main ITIL module is then flattened to TIL. An overview of these steps is shown in Figure 5.

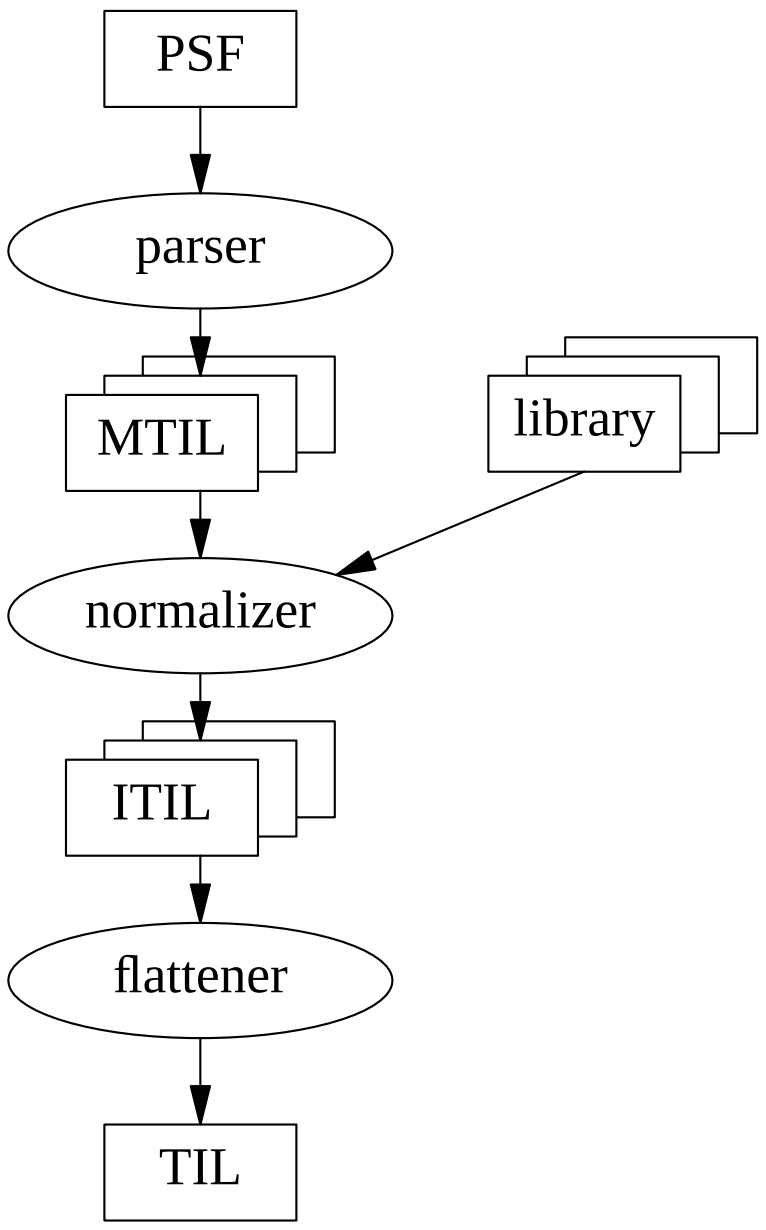

**Figure 5.** Translation from PSF to TIL

The implementation of the PSF compiler is build up from several independent components, controlled by a driver. The compiler driver consist of the following phases.

1. **collecting modules**
   The modules are collected from the files given to the compiler, and missing imported modules are searched for in the libraries

2. **sorting modules**
   The modules are sorted according to their import relation.

3. **splitting files**
   Files scanned in phase 1 that contain more than one module are splitted into files containing one module each.

4. **parsing** (from PSF to MTIL)
   All modules that are out of date, that is the destination file does not exist, or the source file (with extension .psf) is newer than the destination file (with extension .mtil), are parsed.

5. **normalizing** (from MTIL to ITIL)
   All modules that are out of date, that is the destination file does not exist, or the source file (with extension .mtil) is newer than the destination file (with extension .itil) or one of its imported modules (ITIL) is newer, are normalized.

6. **flattening** (from ITIL to TIL)
   The main module is translated from ITIL to TIL.

7. **converting sorts to sets**
   The simulator preprocessor is invoked for converting sorts to sets so that the simulator can deal with them.

8. **checking TRS**
   The term rewrite system checker is invoked.

*3.2 Specification of the compiler*

The complete specification of the compiler will not be displayed, but only those parts that are of interest for turning the compiler into a ToolBus application. The generated animation of the compiler is shown in Figure 6.

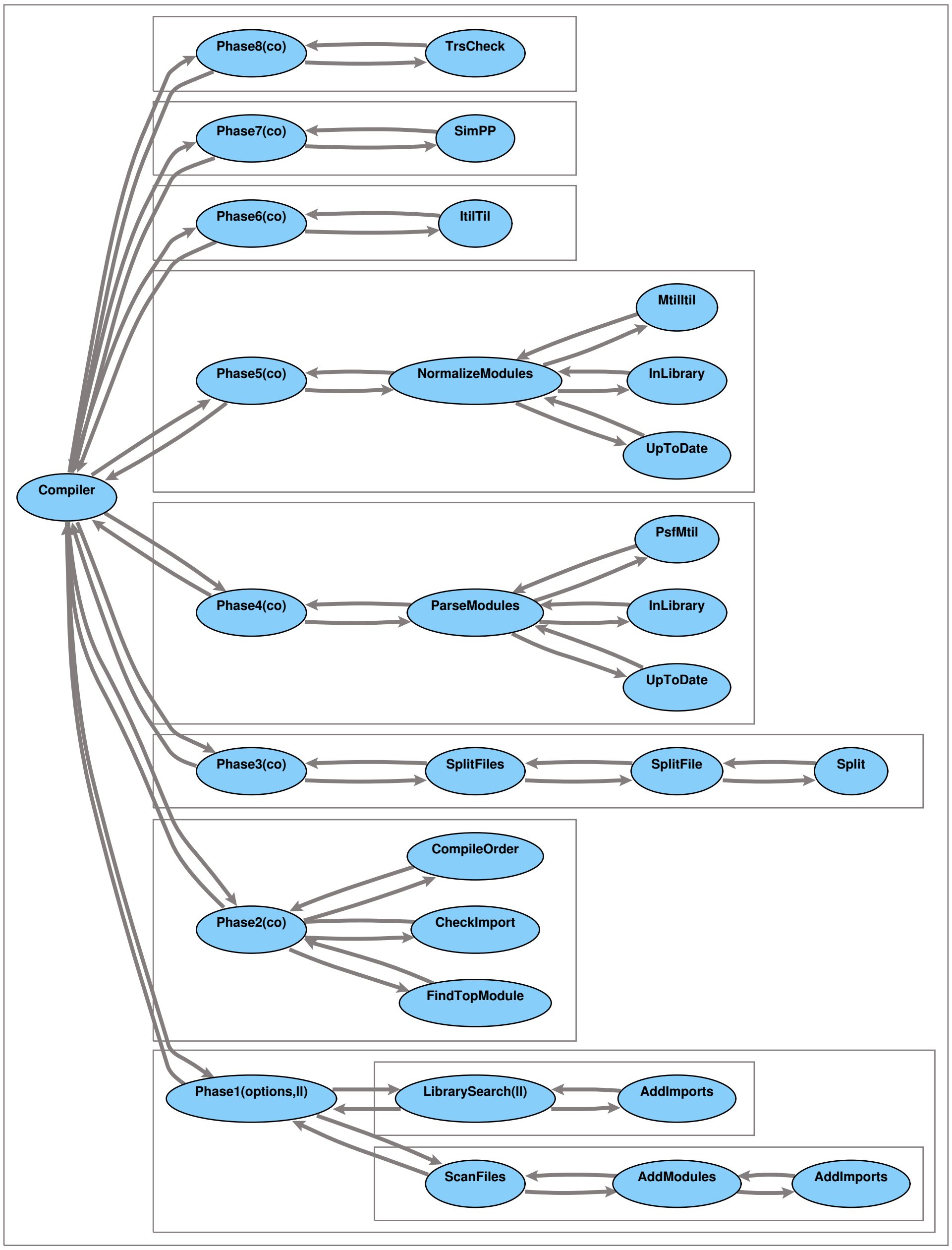

**Figure 6.** Generated animation of the compiler

The processes PsfMtil, MtilItil, ItilTil, SimPP, and TrsCheck are implemented as calls to separate programs. These are used as components and an abstraction is made from their internal workings in the context of this specification.

Just to give some insight in the complexity of the specification, the import structure of the modules of the specification is shown in Figure 7.

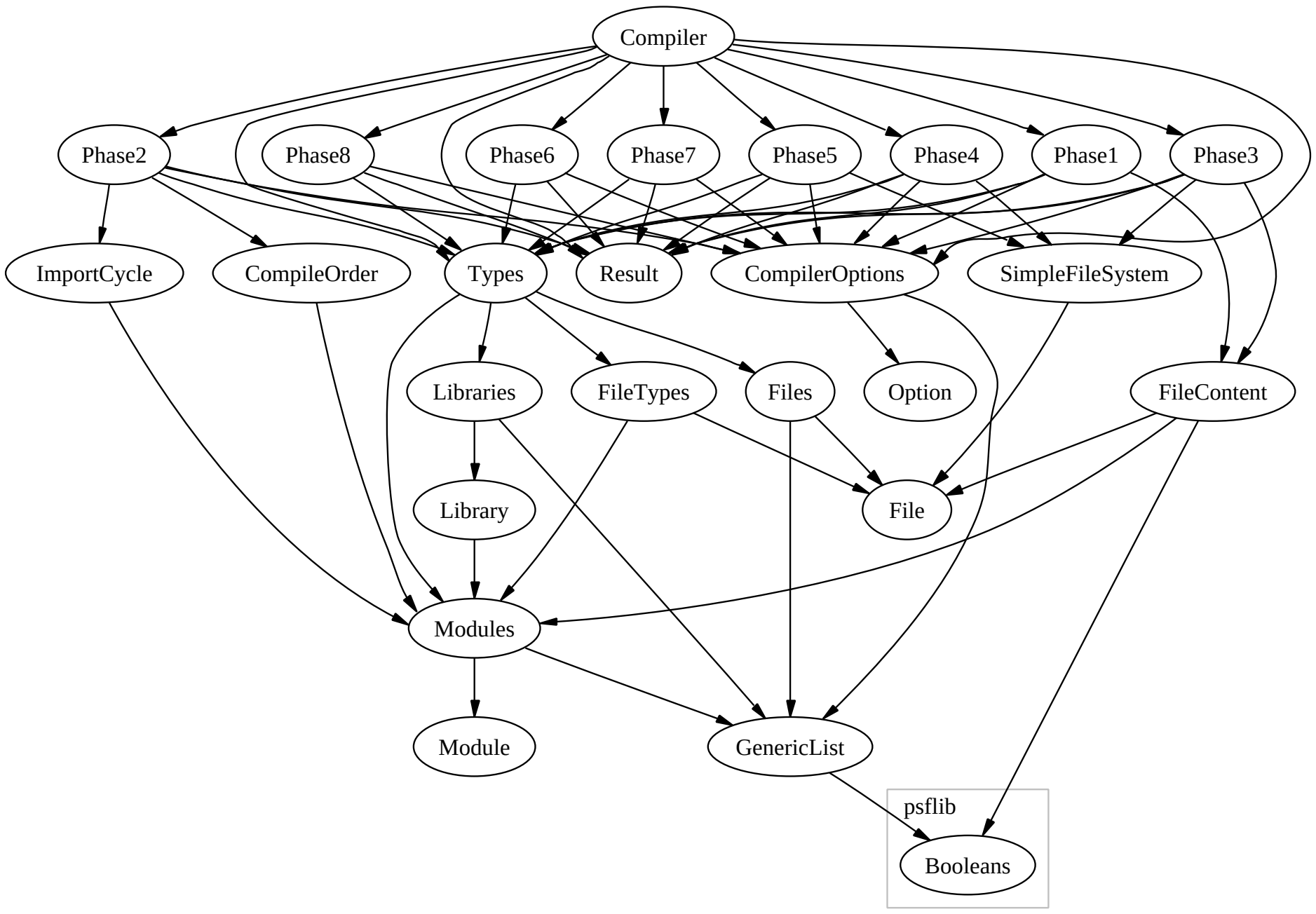

**Figure 7.** Import graph of the specification of the compiler

### *3.3 Specification of the compiler as a ToolBus application*

Instead of calling the parser (process PsfMtil) and normalizer (process MtilItil) directly, they should be called via the ToolBus. This can be accomplished by specifying an adapter for the compiler, and a ToolBus script consisting of the ToolBus processes for the compiler, parser and normalizer. The resulting animation is shown in Figure 8.

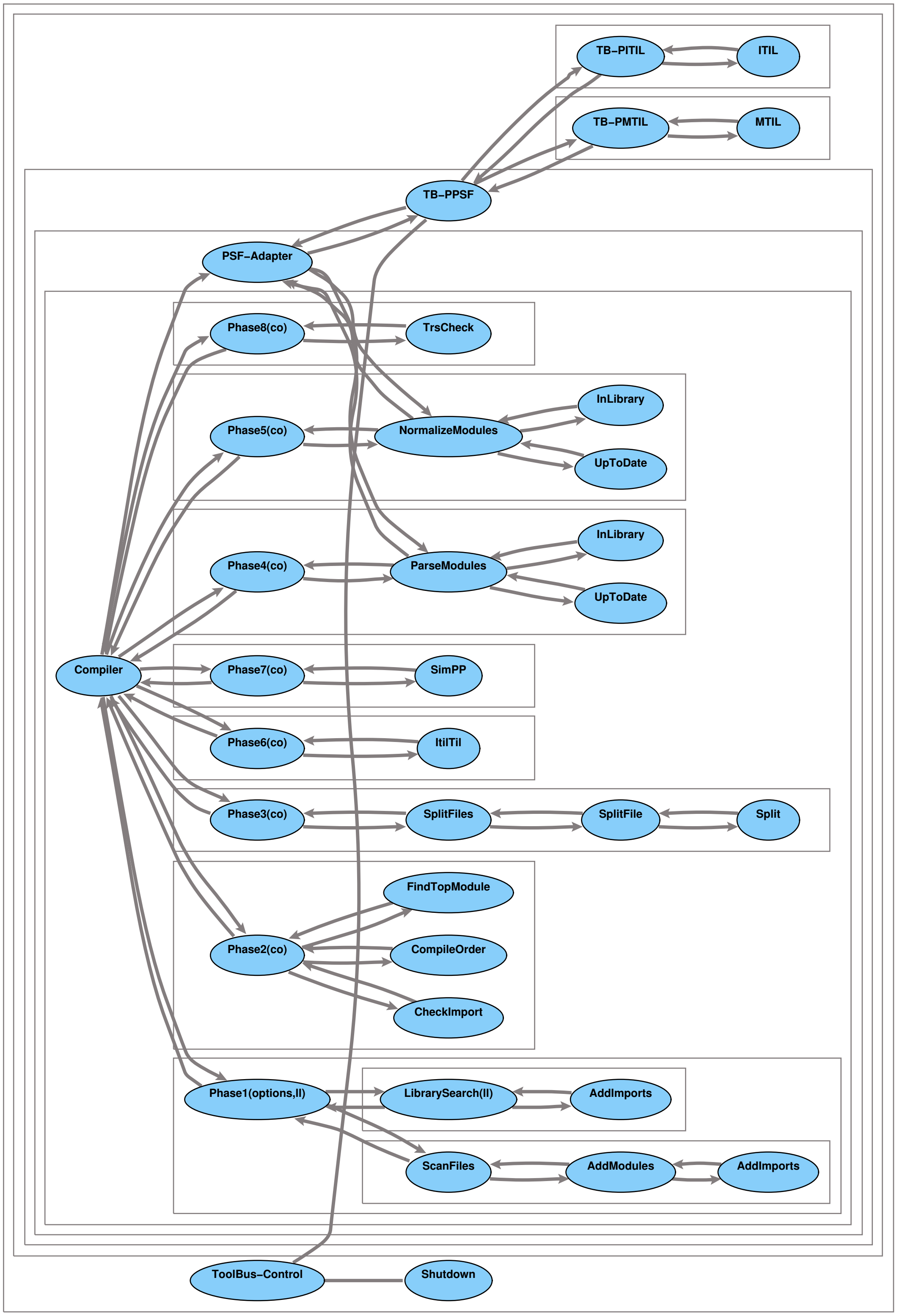

**Figure 8.** Generated animation of the compiler as ToolBus application

The specification of the ToolBus processes is as follows.

**process module** `PPSF`
**begin**

```
    exports
    begin
        processes
            PPSF
    end
    imports
        AdapterPSF,
        ToolFunctions,
        ToolBusPrimitives,
        ToolBusFunctions,
        ToolBus-ID
    definitions
        PPSF =
                AdapterPSF
            ||
                (
                    (
                        sum(args in TBterm,
                            tb-rec-event(PSF, tbterm(tterm(tool-mtil)), args) .
                            tb-snd-ack-event(PSF, tbterm(tterm(tool-mtil))) .
                            tb-snd-msg(psf, mtil, args)
                        ) .
                        sum(result in TBterm,
                            tb-rec-msg(mtil, psf, result) .
                            tb-snd-do(PSF, result)
                        )
                    +   sum(args in TBterm,
                            tb-rec-event(PSF, tbterm(tterm(tool-itil)), args) .
                            tb-snd-ack-event(PSF, tbterm(tterm(tool-itil))) .
                            tb-snd-msg(psf, itil, args)
                        ) .
                        sum(result in TBterm,
                            tb-rec-msg(itil, psf, result) .
                            tb-snd-do(PSF, result)
                        )
                    +   tb-rec-event(PSF, tbterm(quit)) .
                        tb-shutdown
                    ) * delta
                )
end PPSF

process module PMTIL
begin
    exports
    begin
        processes
            PMTIL
    end
    imports
        MTIL,
        ToolBusPrimitives,
        ToolBus-ID
    definitions
        PMTIL =
                MTIL
            ||
                (
                    sum(args in TBterm,
                        tb-rec-msg(psf, mtil, args) .
                        tb-snd-eval(MTIL, args) .
                        sum(result in TBterm,
                            tb-rec-value(MTIL, result) .
                            tb-snd-msg(mtil, psf, result)
                        )
                    ) * delta
                )
end PMTIL

process module PITIL
begin
    exports
    begin
```

```
        processes
            PITIL
    end
    imports
        ITIL,
        ToolBusPrimitives,
        ToolBus-ID
    definitions
        PITIL =
                ITIL
            ||
                (
                    sum(args in TBterm,
                        tb-rec-msg(psf, itil, args) .
                        tb-snd-eval(ITIL, args) .
                        sum(result in TBterm,
                            tb-rec-value(ITIL, result) .
                            tb-snd-msg(itil, psf, result)
                        )
                    ) * delta
                )
end PITIL
```

The import graph of the specification of the compiler as ToolBus application is shown in Figure 9.

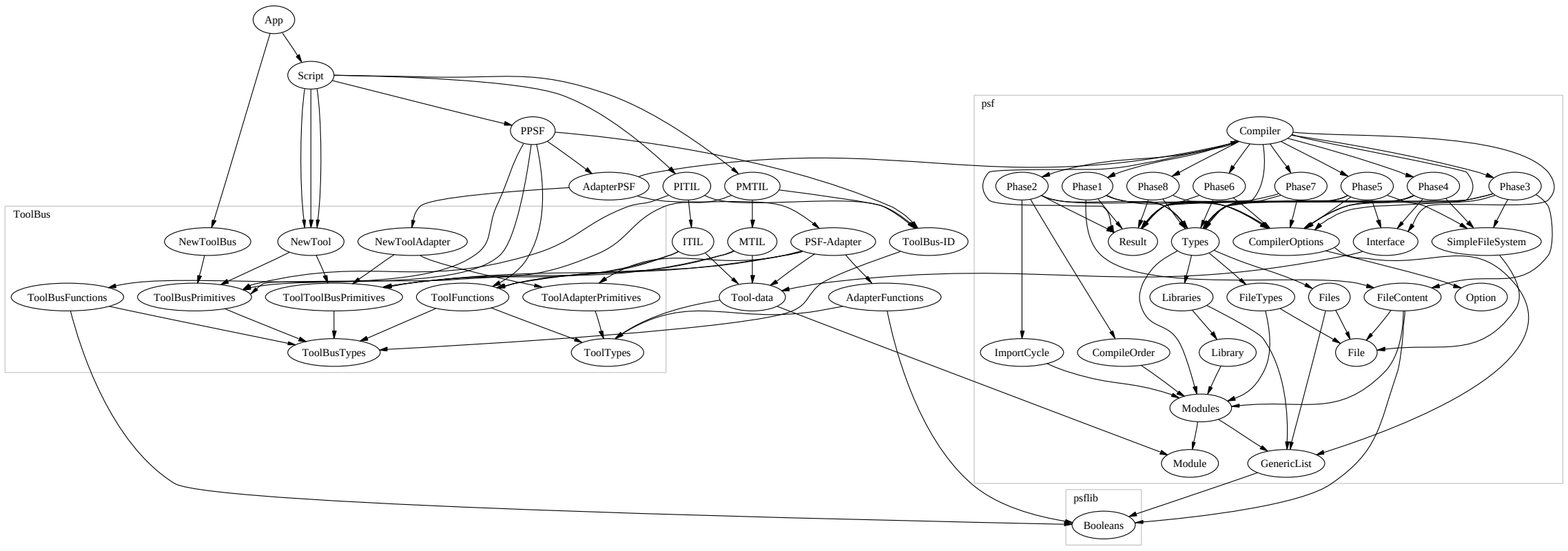

**Figure 9.** Import graph of the specification of the compiler as ToolBus application

### *3.4 Implementation of the compiler as a ToolBus application*

The original implementation of the compiler has been provided with an interface that communicates with the adapter. The adapter is written in Perl [19] as an extension of the Perl-adapter provided with the ToolBus. The ToolBus script is derived from the specification of the ToolBus processes. The parser and normalizer are wrapped with Perl scripts that take care of fetching the exit status of the two tools and sending this information back as a result.
The actual application is a Perl script that provides an environment with all the right settings and invokes the ToolBus, according to the arguments given on the command line.

Although it is not of central interest at this stage, a comparison the performance of the compiler which uses the ToolBus (tbpsf) compared to the original compiler (psf) is given. The tests consists of a complete compilation of the specification of the compiler as a ToolBus application consisting of 49 modules and an update in which only several modules have to be (re)compiled. The tests have been performed on two different machines, one with only one cpu (M1), and one with four cpu's (M4). The timings[4] shown in Table 1 are averages over several runs.

---

4. The configurations were running in normal operation mode, which means that timings are influenced by other processes and load on the file server, and for that reson are very rough.

**Table 1.** Performance of the compilers

| | M1 | | M4 | |
|---|---|---|---|---|
| | complete | update | complete | update |
| psf | 5.5s | 3.0s | 5.5s | 2.8s |
| tbpsf | 17.2s | 5.8s | 7.5s | 3.3s |

It clearly shows that the use of the ToolBus imposes a lot of overhead, largely due to context switching. Because of the four cpu's, the configuration M4 needs fewer context switches, and so has less overhead.

## 4. Software architecture

A software design consist of several levels, each lower one refining the design on the higher level. The highest level is often referred to as the architecture, the organization of the system as a collection of interacting components. In conventional software engineering processes, the architecture is usually described rather informally by means of a boxes-and-lines diagram. Following a lot of research going on in this area architectural descriptions are becoming more formal, especially due to the introduction of architectural description languages (ADL's). A specification in an ADL can be refined (in several steps) to a design from which an implementation of the system can be built. Here, the reverse has to be done. Given a specification of a design in PSF one tries to extract the underlying architecture by means of an appropriate abstraction. The specification of the architecture will still be in PSF, however such that one may generate an animation. This corresponds to the boxes-and-lines diagram but it is fully specified.

In the following sections we describe the possibilities for abstraction, andy apply these to extract the architecture of the compiler.

### *4.1 Abstraction*

In [17], action refinement is used as a technique for mapping abstract actions onto concrete processes, called virtual implementation, which is more fully described in [18]. For extracting the architecture from a specification we use the reverse of action refinement: action abstraction. One may do this by hiding internal actions of a component, and applying process algebra rules to combine consecutive internal actions into a single (internal) action. But also in this transformation step one has to abstract from implementation decisions that do not belong at the resulting higher abstract level. Often this can be done by only looking at the external behavior of a component, its interface.

With parameterized actions, data terms are available which can also be refined. At a certain abstract level one may not care how data is implemented as long as the data is of a particular type. For instance in a message passing system one may deal with any message as just a message without knowing its content. Then for the specification at an abstract level one may use the zero-adic function *message* for the parameter of an action. In the specification at a lower abstraction level this constant can be refined to a more complex term. Data abstraction is the reverse of this, we then replace complex terms with zero-adic functions. With such an abstraction, a receiving action of such a term can now use this zero-adic function instead of a variable coming from a summation construction.

### *4.2 Architecture of the compiler*

In the specification of the compiler the order of compilation steps is laid down. First all modules are parsed and then all modules are normalized. This is an implementation decision. A module can be normalized as soon as it has been parsed and all the modules it imports have been normalized. To abstract from this decision we specify the compiler with the following process.

```
PSF′ =
    skip .
    (
        (
            skip .
```

```
        snd(do(tterm(tool-mtil), tterm(args))) .
        rec(result)
      + skip .
        snd(do(tterm(tool-itil), tterm(args))) .
        rec(result)
      ) * snd(quit)
    )
```

Here, we use the abstract data terms ’args’ and ’result’. This process describes the external behavior of the compiler. The skip actions are abstractions of internal actions.

The adapter for the compiler is defined as follows, where also the abstract form of the data terms are used.

```
PSF-Adapter =
  (
    tooladapter-rec(do(tterm(tool-mtil), tterm(args))) .
    tooltb-snd-event(tbterm(tterm(tool-mtil)), tbterm(tterm(args))) .
    tooltb-rec-ack-event(tbterm(tterm(tool-mtil))) .
    tooltb-rec(tbterm(tterm(result)) .
    tooladapter(tterm(result))
  + tooladapter-rec(do(tterm(tool-itil), tterm(args))) .
    tooltb-snd-event(tbterm(tterm(tool-itil)), tbterm(tterm(args))) .
    tooltb-rec-ack-event(tbterm(tterm(tool-itil))) .
    tooltb-rec(tbterm(tterm(result)) .
    tooladapter(tterm(result))
  ) *
  tooladapter-rec(quit) .
  tooltb-snd-event(tbterm(quit))
```

The parallel composition of PSF’ and PSF-Adapter combined with encapsulation of the communication actions is equivalent to the following process.

```
AdapterPSF’ =
  skip .
  (
    (
      skip .
      tooladapter-comm(do(tterm(tool-mtil), tterm(args))) .
      tooltb-snd-event(PSF, tbterm(tterm(tool-mtil)), args)
      tooltb-rec-ack-event(tbterm(tterm(tool-mtil))) .
      tooltb-rec(result) .
      adaptertool-comm(tterm(result))
    + skip .
      tooladapter-comm(do(tterm(tool-itil), tterm(args))) .
      tooltb-snd-event(PSF, tbterm(tterm(tool-mtil)), args)
      tooltb-rec-ack-event(tbterm(tterm(tool-mtil))) .
      tooltb-rec(result) .
      adaptertool-comm(tterm(result))
    ) *
    tooladapter-comm(quit) .
    tooltb-snd-event(tbterm(quit))
  )
```

We hide all internal actions of this process and replace the data terms with a more abstract form.

```
AdapterPSF’’ =
  skip .
  (
    (
      skip .
      skip .
      tooltb-snd-event(PSF, tool-mtil, args)
      tooltb-rec-ack-event(tool-mtil) .
      tooltb-rec(result) .
      skip
    + skip .
      skip .
      tooltb-snd-event(PSF, tool-mtil, args)
      tooltb-rec-ack-event(tool-mtil) .
      tooltb-rec(result) .
      skip
    ) *
    skip .
```

```
        tooltb-snd-event(quit)
    )
```

The ToolBus process PPSF with the data terms can be written in an abstract form as follows.

```
PPSF' =
        AdapterPSF''
    ||  (
          (
            tb-rec-event(PSF, tool-mtil, args) .
            tb-snd-ack-event(PSF, tool-mtil) .
            tb-snd-msg(psf, mtil, args) .
            tb-rec-msg(mtil, psf, result) .
            tb-snd-do(PSF, result)
          + tb-rec-event(PSF, tool-itil, args) .
            tb-snd-ack-event(PSF, tool-itil) .
            tb-snd-msg(psf, itil, args) .
            tb-rec-msg(itil, psf, result) .
            tb-snd-do(PSF, result)
          ) *
          tb-rec-event(PSF, quit) .
          tb-shutdown
        )
```

After encapsulation of the communication actions between the tool and its ToolBus process this is equivalent to the following.

```
PPSF'' =
    skip .
    (
      (
        skip .
        skip .
        tb-comm-event(PSF, tool-mtil, args) .
        tb-comm-ack-event(PSF, tool-mtil) .
        tb-snd-msg(psf, mtil, args) .
        tb-rec-msg(mtil, psf, result) .
        tb-comm-do(PSF, result) .
        skip
      + skip .
        skip .
        tb-comm-event(PSF, tool-itil, args) .
        tb-comm-ack-event(PSF, tool-itil) .
        tb-snd-msg(psf, itil, args) .
        tb-rec-msg(itil, psf, result) .
        tb-comm-do(PSF, result) .
        skip
      ) *
      skip .
      tb-comm-event(PSF, quit) .
      tb-shutdown
    )
```

Hiding all communications between the tool and the ToolBus process the following result is obtained.

```
PPSF''' =
    skip .
    (
      (
        skip .
        skip .
        skip .
        skip .
        tb-snd-msg(psf, mtil, args) .
        tb-rec-msg(mtil, psf, result) .
        skip .
        skip
      + skip .
        skip .
        skip .
        skip .
        tb-snd-msg(psf, itil, args) .
        tb-rec-msg(itil, psf, result) .
```

```
            skip .
            skip
         ) *
         skip .
         skip .
         tb-shutdown
      )
```

Applying the $\tau$-law $x.\,\tau.\,y = x.\,y$ of our process algebra yields

```
   PPSF'''' =
      skip .
      (
         (
            skip .
            tb-snd-msg(psf, mtil, args) .
            tb-rec-msg(mtil, psf, result)
         +  skip .
            tb-snd-msg(psf, itil, args) .
            tb-rec-msg(itil, psf, result)
         ) *
         skip .
         tb-shutdown
      )
```

The same is done for the processes PMTIL and PITIL.

```
   PMTIL'''' =
      (
         tb-rec-msg(psf, mtil, args) .
         tb-snd-msg(mtil, psf, result)
      ) * delta
   PITIL'''' =
      (
         tb-rec-msg(psf, itil, args) .
         tb-snd-msg(itil, psf, result)
      ) * delta
```

The parallel composition of the above three processes describes the intended architecture. An animation of this architecture is shown in Figure 10.

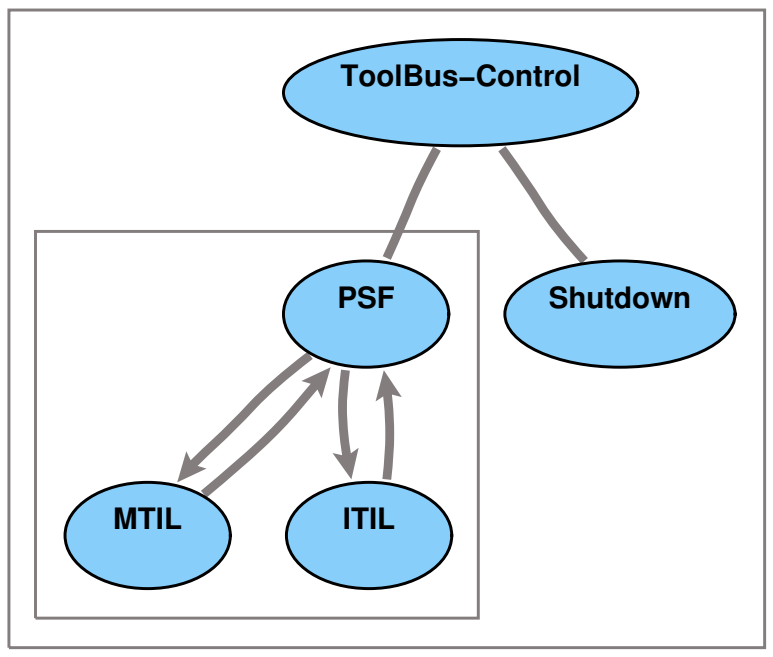

**Figure 10.** Animation of the architecture

With some renaming this can be written in a more suitable form.

```
   Compiler = PSF || MTIL || ITIL
   PSF =
      skip .
      (
         (
            skip .
            snd(psf, mtil, args) .
            rec(mtil, psf, result)
         +  skip .
            snd(psf, itil, args) .
            rec(itil, psf, result)
         ) *
```

```
        skip .
        shutdown
    )
MTIL =
    (
        rec(psf, mtil, args) .
        snd(mtil, psf, result)
    ) * delta
ITIL =
    (
        rec(psf, itil, args) .
        snd(itil, psf, result)
    ) * delta
```

The above PSF text provides a specification of the compiler architecture. The architecture does not enforce any restrictions on the type of connections used to *glue* the various components together. Both the original compiler as well as the reengineered version compiler that makes use of the ToolBus are implementations of this architecture.

## 5. Parallel compiler

The parsing and normalization of modules allows for parallelization. Parsing of modules and the normalization of other modules which already have been parsed and for which all the modules that they import have already been normalized, can be done in parallel.

We build a parallel compiler and reuse as much as possible from the specifications and implementation of the reengineered compiler.

### *5.1 Architecture*

Instead of issuing commands for parsing and normalization of modules, the parallel compiler should compose an information structure that tells which modules have to be parsed and/or normalized and on which modules they depend that also have to be parsed and/or normalized. The compiler has to send this structure to a scheduler which decides when modules are to be parsed or normalized.

We give here the specification of the architecture for the parallel compiler.

```
Compiler = PSF || Scheduler || MTIL || ITIL
PSF =
    skip .
    (
        skip .
        tb-snd-msg(psf, scheduler, compile-info) .
        tb-rec-msg(scheduler, psf, result) .
        tb-shutdown
    +   skip .
        tb-shutdown
    )
Scheduler =
    tb-rec-msg(psf, scheduler, compile-info) .
    (
        (
            skip .
            tb-snd-msg(scheduler, mtil, args)
        +   tb-rec-msg(mtil, scheduler, result)
        +   skip .
            tb-snd-msg(scheduler, itil, args)
        +   tb-rec-msg(itil, scheduler, result)
        ) * tb-snd-msg(scheduler, psf, result)
    )
MTIL =
    (
        tb-rec-msg(scheduler, mtil, args) .
        tb-snd-msg(mtil, scheduler, result)
    ) * delta
ITIL =
```

```
(
   tb-rec-msg(scheduler, itil, args) .
   tb-snd-msg(itil, scheduler, result)
) * delta
```

An animation of this architecture is shown in Figure 11.

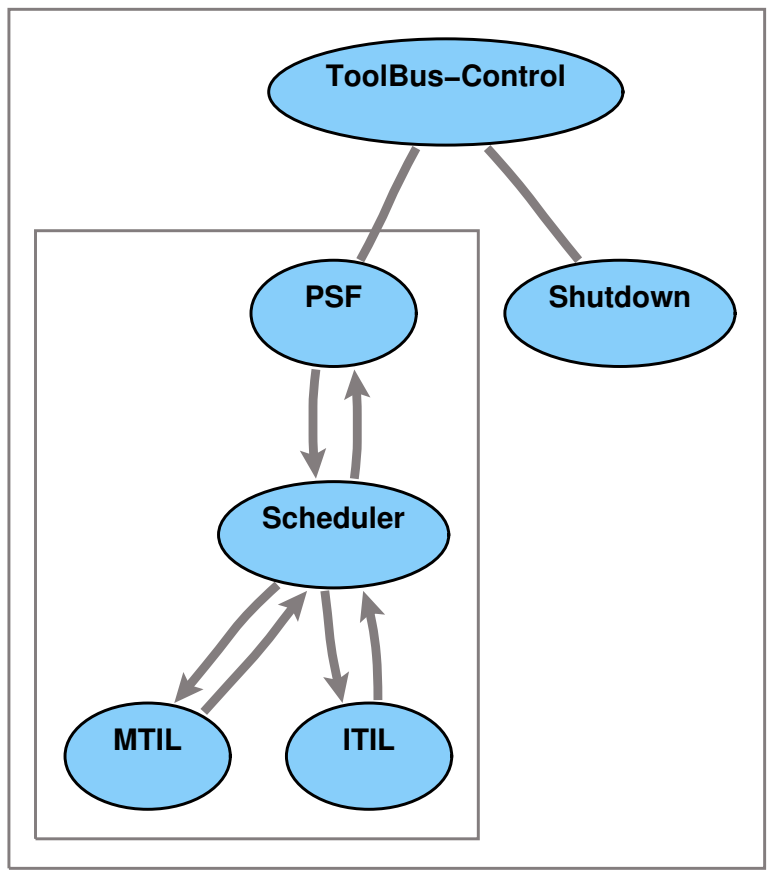

**Figure 11.** Animation of the architecture

This specification features only one MTIL and one ITIL process, but this scheme allows for more MTIL and ITIL processes in parallel. Whichever process is free can pick up a request from the scheduler to parse or normalize a module.

Although the specification of the architecture contains separate processes for compiler and scheduler, it does not imply that these need to be implemented as separate tools. The scheduler can be incorporated in the compiler, as we show below.

The parallel composition of PSF and Scheduler, is equivalent to the following process.

```
skip . (
   skip . tb-comm-msg(psf, scheduler, compile-info) .
   (
      P * tb-comm-msg(scheduler, psf, result) . tb-shutdown
   )
+  Q
)
```

Here, `P` stands for the alternative composition of the send and receive actions in the Scheduler process, and `Q` stands for **skip .** `tb-shutdown`.

Hiding the communications between compiler and scheduler results in the following.

```
skip . (skip . skip . (P * Q) + Q)
```

Applying the $\tau$-law $x.\tau.y = x.y$ gives

```
skip . (skip . (P * Q) + Q)
```

Applying the rule for iteration $x * y = x.(x * y) + y$ gives

```
skip . (skip . (P . (P * Q) + Q) + Q)
```

Applying the $\tau$-law $x.(y + \tau.z) = x.(y + \tau.z) + x.z$ in reverse gives

```
skip . skip . (P . (P * Q) + Q)
```

Applying the rule for iteration in reverse and the $\tau$-law $x.\tau.y = x.y$ gives

```
skip . (P * Q)
```

Replacing `P` and `Q` gives us

```
skip .
```

```
(
     skip .
     tb-snd-msg(scheduler, mtil, args)
  +  tb-rec-msg(mtil, scheduler, result)
  +  skip .
     tb-snd-msg(scheduler, itil, args)
  +  tb-rec-msg(itil, scheduler, result)
  ) * skip .
  tb-shutdown
)
```

This looks the same as the compiler process in the architecture of tbpsf but then with the sending and receiving actions in parallel with the scheduler process in this architecture.

### *5.2 Specification of the parallel compiler*

As we already mentioned in the previous section, there are several options for the cooperation of the compiler and the scheduler. A possiblity is to incorporate the scheduler in the compiler and let the scheduler part take care of the connections with the ToolBus. Here, however, we have chosen to implement the Scheduler as separate process (tool) to be connected to the ToolBus which gets its information from the compiler over the ToolBus.

We reuse a large part of the specification of the compiler for the specification of the parallel compiler. The parsing and normalization phases are replaced by a phase that builds up a Compiler-Information structure. The ToolBus processes are adjusted and extended to reflect the processes in the specification of the architecture. And the specification of the scheduler is added.

The animation of the parallel compiler is shown in Figure 12.

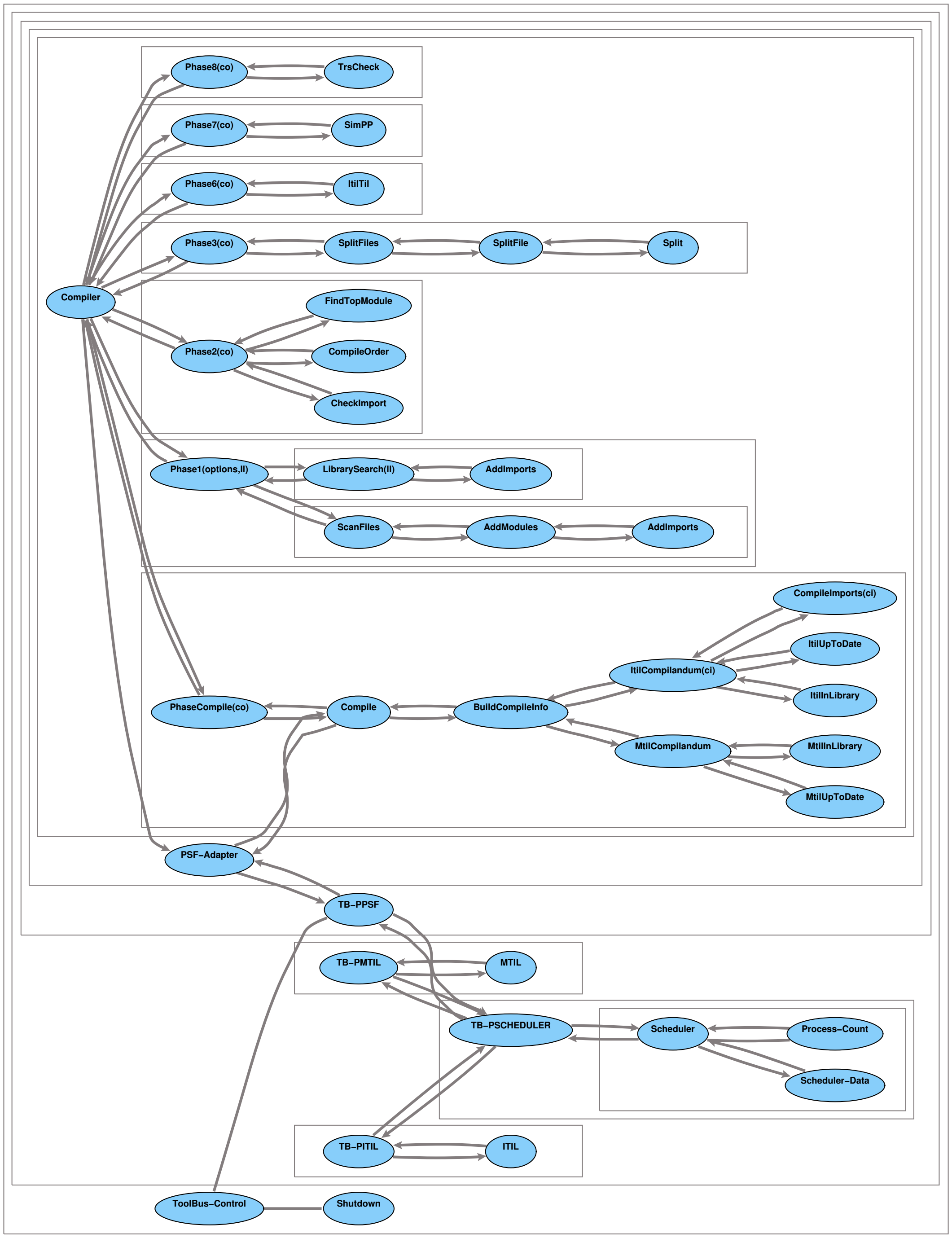

**Figure 12.** Generated animation of the parallel compiler as ToolBus application

The import graph for the specification of the parallel compiler is shown in Figure 13.

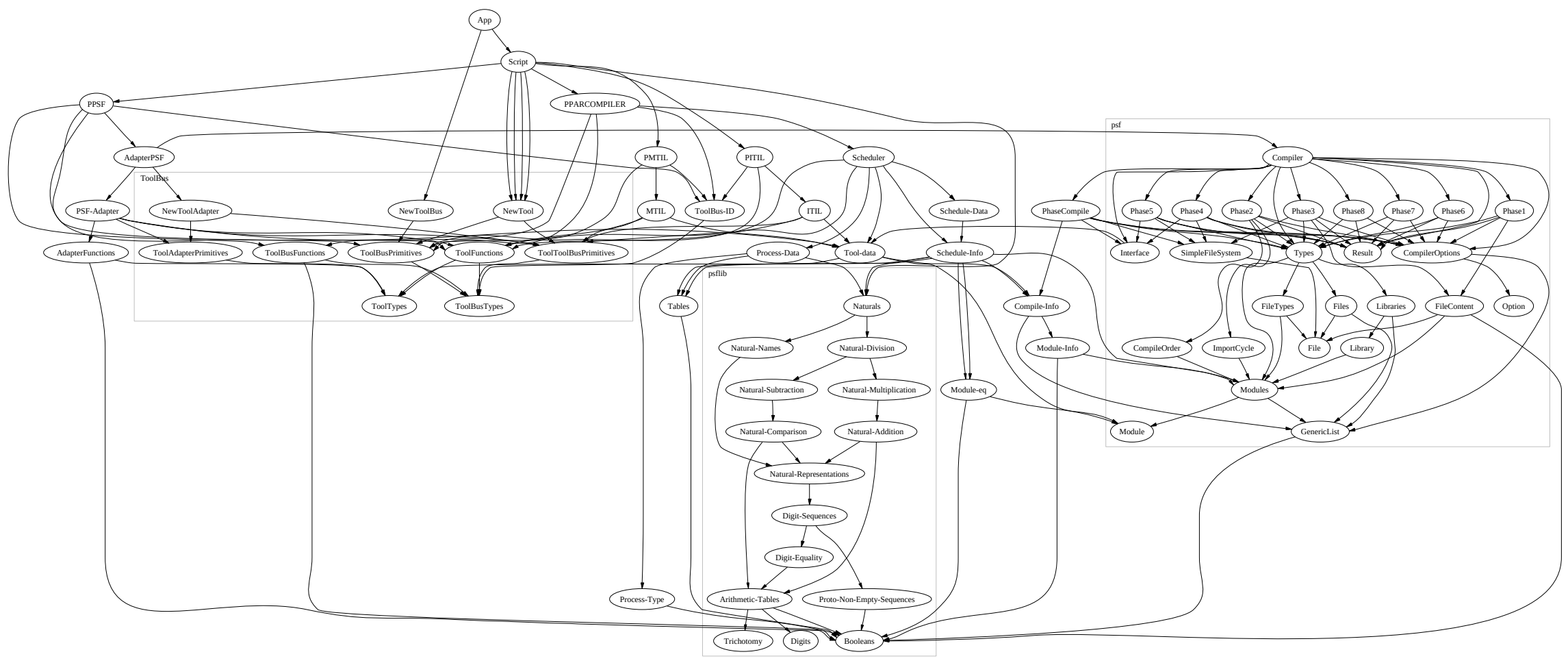

**Figure 13.** Import graph of the specification of the parallel compiler

The module Naturals and all its imports and the module Tables stem from the standard library of PSF. Naturals is used for counting the MTIL and ITIL processes that have to be started in the specification of the ToolBus script and Tables is used for the construction of the Compiler-Information structure.

### *5.3 Implementation of the parallel compiler*

The implementation of the compiler has been extended with a phase for building the Compiler-Information structure which can be invoked instead of the parsing and normalizing phases, controlled by an option. The scheduler has been implemented in Perl. The actual application is a Perl script that provides an environment with the right settings and which will invoke the ToolBus according to the arguments given on the command line. This script also gives the possibility to start the parallel compiler with indicated numbers of parsing and normalization processes.

In Table 2 the performance of the parallel compiler is shown for several combinations of numbers of parsing and normalization processes, for the complete compilation of the specification of the compiler as a ToolBus application.

**Table 2.** Performance of the parallel compiler

| # processes | | complete | |
|---|---|---|---|
| mtil | itil | M1 | M4 |
| 1 | 1 | 12.8s | 6.0s |
| 1 | 2 | 11.7s | 5.4s |
| 1 | 3 | 11.6s | 5.4s |
| 2 | 1 | 12.9s | 6.1s |
| 2 | 2 | 12.1s | 5.2s |
| 2 | 3 | 11.9s | 4.8s |
| 2 | 4 | 11.6s | 4.9s |
| 3 | 2 | 11.7s | 5.0s |
| 3 | 3 | 11.6s | 4.8s |
| 3 | 4 | 11.6s | 4.9s |
| 4 | 4 | 11.6s | 4.7s |

We see that on configuration M1 the parallel compiler has a better performance than tbpsf, but it is not faster than psf. So the communication overhead connected with the ToolBus is too large to overcome on this configuration. Parallel compilation on configuration M4 is faster than psf, although not much, because the amount of work that can be done in parallel is limited by the imposed order of compilation of the modules due to their import relation.

## 6. Conclusions

We have made a specification for the compiler in the ToolKit of PSF and a specification of a library of ToolBus internals, which we used for developing a specification of the compiler with the use of this ToolBus library for coordination of the components of the compiler. From this specification, we were able to extract a specification of the architecture of the compiler. Furthermore, we have build a parallel compiler by developing a specification of the architecture, a refined specification, and an implementation, with reuse of as much as possible of the specification and implementation for the compiler as ToolBus application.

PSF turned out to be very useful. Its modularization and parameterization features made the use of a library for the ToolBus internals possible, which makes the specification of a software system with interacting components much easier. Specification can be done at various abstract levels, as we have shown by making specifications of the compiler close to the implementation level as well as at the architectural level. The latter indicates that PSF with the ToolBus library can be used as an ADL. The animation facility coupled with simulation gives a very good view of which processes are involved in certain communications, much more than a visual inspection of the PSF specification itself can provide. The animation of the architecture is very useful for explaining the software system to stakeholders who have not been involved in the software design process.

Although we have used the PSF ToolBus library in our specification, an implementation does not necessarily need to use the ToolBus. All the connections between processes in the ToolBus part of the specification can be implemented in numerous ways. These connections are abstract, and the ToolBus provides an implementation.

In this paper, we have reported on the experience gained through reengineering a compiler that already consisted of separately implemented components, but one should also acquire experience with starting at the software architecture level and working towards an implementation. Here lies the use of the PSF simulator. This is a complex piece of software with integrated graphical user interface implemented in the X Window System. The user interface could well be implemented as separate components in Tcl/Tk. In this way, not only the interface can be changed easily, but also a simulator kernel for a different process algebra notation can be used if that is preferable.